\documentclass[twocolumn,letterpaper,aps,prc,longbibliography,superscriptaddress,showpacs,floatfix]{revtex4-1}

\usepackage{comment}

\usepackage[utf8]{inputenc}
\usepackage{t1enc}
\usepackage{setspace}
\usepackage{indentfirst}
\usepackage{amsmath, amsthm, amssymb}

\usepackage[pdftex]{graphicx}
\usepackage{epstopdf}
\usepackage{subfigure}
\usepackage{appendix}
\usepackage[bottom]{footmisc}
\usepackage{float}
\usepackage{anysize}
\usepackage{underlin}
\usepackage{fancyhdr}
\usepackage{epigraph}
\usepackage{url}
\marginsize{2cm}{2cm}{2cm}{1.5cm} 
\usepackage{tikz}
\graphicspath{{./figures/}}
\usepackage{wrapfig}
\usepackage{rotating}
\usepackage[colorlinks]{hyperref}

\newcommand{\pp}{p+p }

\newcommand{\pbpb}{Pb+Pb }
\newcommand{\sqsntwo}{$\sqrt{s_{\textmd{NN}}}$}
\newcommand{\sNN}{\sqrt{s_{\textmd{NN}}}}
\newcommand{\auau}{Au+Au }

\begin{document}

\title{Resonance production in partial chemical equilibrium}

\author{S\'andor L\"ok\"os}
\affiliation{Institute of Nuclear Physics, Polish Academy of Sciences, Cracow, Poland}
\affiliation{MATE Institute of Technology KRC,Gy\"ongy\"os, Hungary}
\affiliation{E\"otv\"os Lor\'and Tudom\'anyegyetem, Budapest, Hungary}
\author{Boris Tom\'a{\v s}ik}
\affiliation{Fakulta pr\'irodn\'ych vied, Univerzita Mateja Bela, Bansk\'a Bystrica, Slovakia} 
\affiliation{Fakulta jadern\'a a fyzik\'aln\v{e} in\v{z}en\'yrsk\'a, \v{C}esk\'e vysok\'e u\v{c}en\'i technick\'e v Praze, Praha, Czech Republic}

\date{\today}

\begin{abstract}
    In high energy collisions, a dense, strongly interacting medium could be created, the quark gluon plasma. In  rapid expansion, from the soup of quarks and gluons a gas of resonance and stable particles is formed at the chemical freeze-out and after that, as the system cools down, the kinetic freeze-out takes place and interaction between particles ceases. By measuring resonance ratios one could get information about the dominant physical processes in the intermediate temperature ranges, i.e. between the chemical and kinetic freeze-out. These quantities are measured at RHIC and LHC energies. In the present analysis we employ the hadron resonance gas model assuming partial chemical equilibrium to characterize these measured data. We calculate the ratios of several resonances to their stable counterpart and compare these model calculations to available experimental data.
\end{abstract}

\maketitle


\section{Introduction}
\label{s:intro}

Collisions of atomic nuclei at ultrarelativistic energies provide an environment for studying the properties of very hot and dense strongly interacting matter. Hadrons, which escape from the fireball after its breakup carry direct information about its dynamical state at the end of the evolution. 

Resonances mediate the interactions among hadrons. Thus a measurement of their production in nuclear collisions carries information about the interactions that are going on in the hot medium, especially towards the end of its evolution. 

A standard baseline that is used for the interpretation of hadron data is built upon the idea of statistical production of hadrons. It has been shown, that at lowest order of the virial expansion, interactions between ground state hadrons can be incorporated into the statistical model by introducing the resonances into the partition function and treating them as free particles \cite{Dashen:1969ep}. 

Statistical model has been quite successful in describing the abundances of ground state hadrons \cite{Becattini:2005xt,STAR:2008med,ALICE:2013mez,Andronic:2018qqt,Andronic:2017pug,Bhattacharyya:2019wag,Bhattacharyya:2019cer}, and even---which is rather puzzling---clusters like deuterons, tritons, or $^3$He \cite{ALICE:2019fee,Biswas:2020kpu}. It leads to the introduction of the so-called \emph{chemical freeze-out}. This is the thermodynamic state of the fireball, specified by its temperature, chemical potentials, and volume, which reproduces the observed abundances of stable hadrons. It should be stressed that it also accounts for the production of stable hadrons from (chains of) decays of resonances, which are present in the thermalised fireball. The average number of resonances is also set by the same parameters. 

However, transverse momentum spectra seem to indicate hadron production from locally thermalised fireball at much lower temperature \cite{Melo:2015wpa}. The fireball thus cools down in the hadronic phase from the chemical freeze-out down to the thermal freeze-out, while the final state abundances of ground state hadrons must be fixed. 
Note, though, that there are other studies, also, which do not indicate such a low kinetic freeze-out temperature \cite{Mazeliauskas:2019ifr}. 
The issue is thus somewhat inconclusive at the moment. 

In an extended hadron phase, a 
decrease of the temperature affects the ratio of resonance abundances to those of stable hadrons. 
The condition of fixed stable species abundance implies specific prescription for non-equilibrium chemical potentials of individual species \cite{Bebie:1991ij}. 
Such a state is usually described as Partial Chemical Equilibrium (PCE).
Consequently, it also influences the resonance-to-stable ratios of abundances. 
Hence, the measurement of this ratio would probe such a scenario. 
One could argue that the proper treatment of resonance production would be by employing transport
simulation. 
This would also be the relevant treatment for the case that resonances
cannot be reliably measured due to rescattering of their 
decay products \cite{Knospe:2015nva}.
Nevertheless, we want to explore PCE as a simple and economic alternative to the 
complicated and computationally expensive transport simulations. In this study 
we thus investigate the limits of applicability of the PCE model. 

In this paper we therefore
entertain the idea of a scenario with an extended hadronic 
phase and PCE. For this scenario we
calculate the production of resonances 
and determine the resonance-to-stable ratios for selected types of resonances which have been measured experimentally. 
Within the used model, such a ratio can be assigned to a value of the temperature, although this may not always be possible. 
The extracted temperatures can be compared with those of kinetic freeze-out in order to see if the instantaneous freeze-out is a good approximation or to what extent it is distant from reality.

We describe the basics of the statistical model with PCE in Section \ref{s:model}. 
In Section \ref{s:data}, we explain our selection of experimental data and the method of comparison of theoretical results to them. 
The actual results are summarised in Section \ref{s:results}, followed by conclusions in Section \ref{s:conc}.


\section{Description of the model}
\label{s:model}

\subsection{Hadron resonance gas model}

The analysis is performed in  framework of the hadron resonance gas model \cite{Dashen:1969ep}. The model is given by the logarithm of its partition function:
\begin{align}
\nonumber
    \ln Z (T,\mu,V)& =
    \sum_i \ln Z_i(T,\mu_i,V) \\
    & = 
    \sum_i \pm \frac{g_i V}{2\pi^2} \int_0^\infty \, p^2 \ln\left[ 1\pm e^{\frac{\mu_i-\varepsilon(p)}{T}} \right] dp
\end{align}
where the sum runs over the stable and resonance hadron species, $p$ is the momentum, $\varepsilon(p)=\sqrt{p^2+m_i^2}$ is the energy, $g_i$ is the spin degeneracy factor, and the $\pm$ sign corresponds to the Bose/Fermi case respectively. The  chemical potentials $\mu_i$ are set for every particle species. From the partition function, based on thermodynamical identities, one can get the partial pressure, energy density, and number density for species $i$
\begin{align}
    P_i &= \frac{T\ln(Z_i)}{V} \\
    e_i &= \frac{g_i}{2\pi^2}\int_0^{\infty} \frac{p^2 \varepsilon(p) }{\exp\left[ \frac{\varepsilon(p)-\mu_i}{T} \right] \pm 1 }  dp\\
    n_i &= \frac{g_i}{2\pi^2}\int_0^{\infty} \frac{p^2 }{\exp\left[ \frac{\varepsilon(p)-\mu_i}{T} \right] \pm 1 } dp . \label{eq:ni}
\end{align}
In our calculations , we shall also need the entropy density, which can be determined from the thermodynamic relation 
\begin{align}
    s= \sum_i \frac{P_i + e_i -  n_i  \mu_i}{T}
\label{eq:entropy}
\end{align}
where the sum, again, runs over the hadron species including the resonances.


\subsection{Partial chemical equilibrium}

During the time evolution of the fireball, two freeze-out stages take place.
As it cools down and reaches certain temperature, the chemical freeze-out happens when inelastic processes cease. 
Data indicate that this happens in the proximity of the hadronisation transition \cite{Andronic:2017pug}. 
After further cooling, the kinetic freeze-out is reached where all interactions between the particles are assumed to disappear.

Measurements showed that the kinetic and the chemical freeze-out temperature differ by about 50--70 MeV/c$^2$ (see in Ref. \cite{Melo:2019mpn}).
However, the multiplicities are frozen at the chemical freeze-out temperature. 
Consequently, the subsequent cooling and expansion should evolve in such a way that the average \emph{effective} number of the stable particle species is conserved. 
Here, the hadrons which are produced from decays of unstable resonances are also included. This can be formulated as
\begin{align}
\langle N^\textmd{eff}_h \rangle = \sum_i c_{i \rightarrow h} \langle N_i \rangle.
\end{align}
The sums runs over both the stable as well as resonance hadron species and $\langle N_i \rangle$ is the average number of species $i$. The weights $c_{i \rightarrow h}$ mean the average number of hadron $h$ that originates from one resonance $i$. (N.B.: $c_{h \rightarrow h}=1$.)

Since the abundance ratios between stable hadrons are kept constant even in spite of decreasing temperature, this is a non-equilibrium feature which will be parametrised with the help of chemical potentials.
However, PCE also  means that the resonances are in equilibrium with their daughter particles. 
Consequently, chemical potential of  resonance species is given by the sum of chemical potentials of its daughters.  
A simple example is the $\Delta^{++}$ with only one decay channel into a proton and pion
\begin{align*}
    \Delta^{++} &\rightarrow p + \pi^{+} \\
    \mu_{\Delta^{++}} &= \mu_p + \mu_{\pi^{+}}\, . 
\end{align*}
Resonances with more then one decay channel are also considered. 
Their chemical potential is then obtained as weighted average with branching ratios as weights. Let us use $\Delta^{+}$ as an example
\begin{align*}
    \Delta^{+} &\rightarrow n + \pi^{+} \hspace{0.5cm}\textmd{or}\hspace{0.5cm} 
    \Delta^{+} \rightarrow p + \pi^{0} \\
    \mu_{\Delta^{+}} &= a_{[\Delta^{+} \rightarrow n 
     + \pi^{+}]} ( \mu_n + \mu_{\pi^{+}} ) \\
    & \qquad  + a_{[\Delta^{+} \rightarrow p + \pi^{0}]} ( \mu_p + \mu_{\pi^{0}} )\, . 
\end{align*}
where the numbers $a_{[...]}$ denote the branching ratios.
After summing up contributions from chain decays of heavier resonances, 
the chemical potential of resonance species $i$ is given as
\begin{align}
    \mu_i = \sum_h c_{i\rightarrow h}\mu_h\,  ,
\end{align}
where  the sum runs through all \emph{stable} hadrons species.

As we mentioned above, the effective numbers of stable hadrons are conserved.
This requires that each stable species obtain their own $\mu_h$ as the system cools down. 
However, chemical potentials as functions of temperature cannot be calculated from the condition of constant $\langle N^\textmd{eff}_h \rangle$, solely, because the numbers also depend on volume, which is unknown and also changes. The trick is in the assumption of isentropic expansion and the use of entropy $S$ as another conserved quantity. 
The volume-independent ratio $\langle N^\textmd{eff}_h \rangle / S $ is then also conserved. 
We can thus work with the corresponding densities
\begin{align}
    \frac{\langle N^\textmd{eff}_h \rangle}{S} & = 
    \frac{n^\textmd{eff}_h}{s} \\
    n^\textmd{eff}_h &= \sum_i c_{i\rightarrow h} n_i\,  ,
    \label{eq:neff}
\end{align}
where $s$ can be determined from eq.~(\ref{eq:entropy}) and $n_i$'s from eq.~(\ref{eq:ni}). 
Hence, the following system of algebraic equations can be used for the calculation of $\mu_h(T)$'s
\begin{align}
    \frac{n^\textmd{eff}_h(T) }{s(T)} = \left.\frac{n^\textmd{eff}_h(T)}{s(T)} \right|_{T=T_\textmd{chem}}
    \label{eq:evolvemu}
\end{align}
where $T_\textmd{chem}$ is the temperature of the chemical freeze-out and the equations are indexed by $h$.

The starting point of the evolution is given at the chemical freeze-out.
Chemical equilibrium with the temperature ($T_\textmd{chem}$),  baryochemical potential ($\mu_B$), the strangeness chemical potential ($\mu_S$) is assumed so that for each species its chemical potential is given as 
\begin{align}
    \mu_i = B_i\mu_B + S_i\mu_S\,  .
\end{align}
Strange species may be undersaturated and this is parametrised by fugacity factor $\gamma_S$.
The partition function then becomes
\begin{align}
    \ln Z (T,\mu,V)= \sum_i \pm \frac{g_i V}{2\pi^2} \int_0^\infty dp p^2 \ln\left[ 1\pm \gamma_S^{|S_i|}e^{\frac{\mu_i-\varepsilon(p)}{T}} \right].
\end{align}
Hence, the initial values for the non-equilibrium chemical potentials are determined as 
\begin{align}
    \mu_i(T=T_\textmd{chem}) = B_i\mu_B + S_i\mu_S + |S_i| T_\textmd{chem}
    \ln \gamma_S\,  .
\end{align}


\subsection{Ratios of abundancies}

Our goal is to calculate the multiplicity ratios of resonance species to stable hadrons. 
The multiplicities scale with the volume, but it drops out in such ratios and it is sufficient to determine the ratios of the \emph{effective} densities. 
Resonance production of the stable hadrons is included as in eq.~(\ref{eq:neff}). 
In the same way, also the effective densities of resonances include contributions from decays of heavier resonances. 

The ratios will be calculated as functions of temperature. 
Through a comparison with experimental data, a temperature will be determined at which the best agreement is reached. 
In a scenario with instantaneous kinetic freeze-out, this should correspond to the freeze-out temperature of the given species.

We considered the following ratios as they were measured in heavy-ion collisions: $\phi/K^-$, $K^*/K^-$, $\rho^0/\pi$, $\Lambda^*/\Lambda$, $\Sigma^*/\Lambda$.


\section{Comparison with data}
\label{s:data}

\subsection{Description of the analysed data}

The resonance ratios we analyse in this paper were measured by the ALICE experiment in \pbpb collisions at \sqsntwo\ = 2.76 TeV and by the STAR experiment in \auau collisions at \sqsntwo\  = 200 GeV. 
The data sets along with the references to publications of the measured values are listed in 
Table~\ref{tab:alldata}. 
\begin{table*}[t]
    \caption{The data sets considered in the analysis. We will refer to the different data sets by the running number in the first column.}
    \label{tab:alldata}
    \centering
    \begin{tabular}{|c|c|c|c|c|}
        \hline
        \# & particle ratio(s) & Experiment & Energy [GeV] & Ref. \\
        \hline
        1 & $K^*(892)^0/K^-$, $\phi(1020)/K^-$ & ALICE & 2760 & \cite{ALICE:2014jbq} \\
        2 & $2\rho(770)^0/(\pi^+ + \pi^-)$ & ALICE & 2760 & \cite{ALICE:2018qdv} \\
        3 & $\Lambda(1520)/\Lambda$ & ALICE & 2760 & \cite{ALICE:2018ewo} \\
        4 & $K^*(892)^0/K^-$ & ALICE & 2760 & \cite{ALICE:2017ban} \\
        5 & $K^*(892)^0/K^-$, $K^*(892)^0/\phi(1020)$ & STAR & 200 & \cite{STAR:2010avo} \\
        6 & $\phi(1020)/K^-$ & STAR & 200 & \cite{STAR:2008bgi} \\
        7 & $\Sigma(1385)/\Lambda$, $\Lambda(1520)/\Lambda$, $K^*(892)^0/K^-$, $\phi(1020)/K^-$ & STAR & 200 & \cite{STAR:2006vhb} \\
        \hline
    \end{tabular}
\end{table*}
There are other ratios available in the literature measured in different colliding systems, e.g. \pp and p+Pb, but we only use the  heavy-ion data, here.

In the literature, the data are presented depending on centrality, which is in different papers quantified as  $\left( {dN}/{d\eta} \right)^{1/3}$, $N_\textmd{part}$, $\frac{dN}{d\eta}$, $\frac{dN}{dy}$ or centrality percentile. 
In order to put them on equal footing, we chose $\left({dN}/{d\eta} \right)^{1/3}$ as the variable for the analysis and convert the others into this by utilizing the related publications of the given experiment (for ALICE, see \cite{ALICE:2010mlf}; for STAR, see \cite{STAR:2008med,STAR:2021iop}).


\subsection{The comparison method}

\begin{figure}[t]
    \centering
    \includegraphics[width=0.48\textwidth]{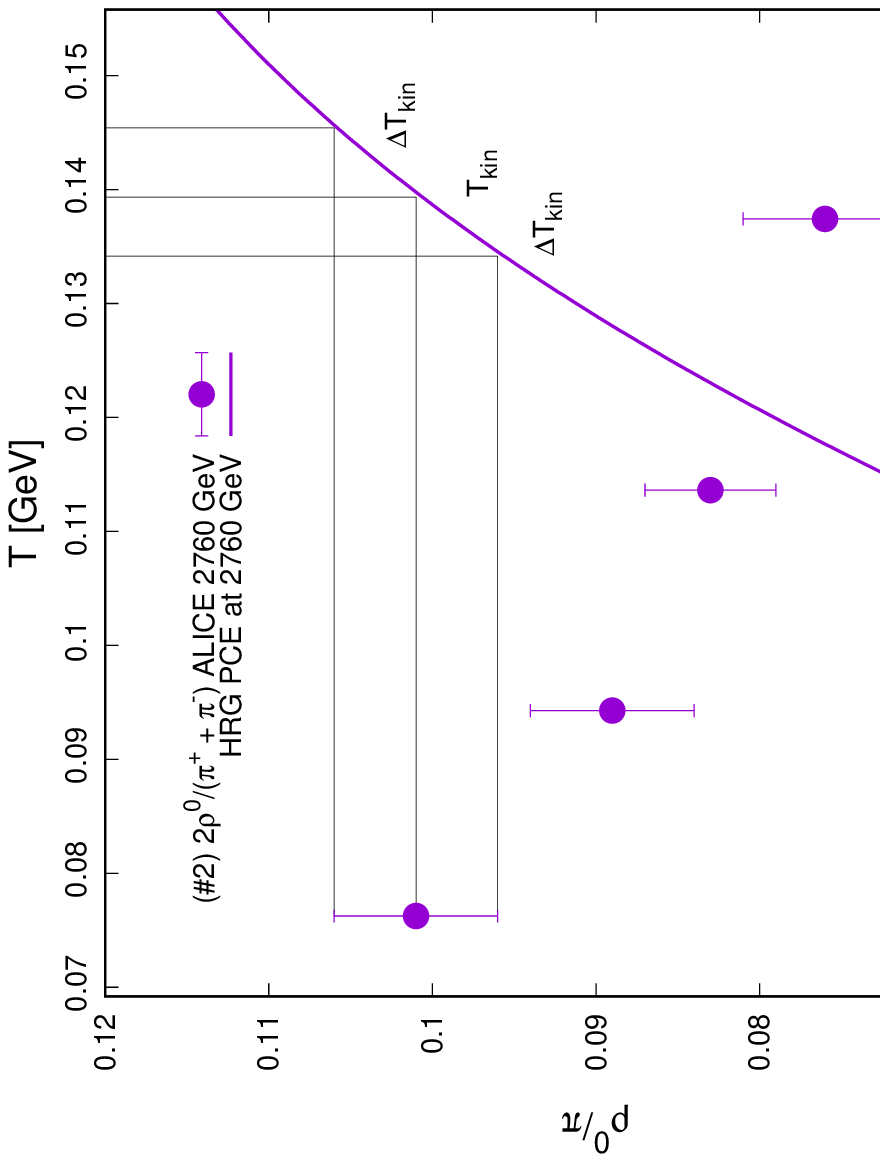}
    \caption{The $\rho^0/\pi$ ratio as function of centrality (experimental data) and temperature (theoretical curves). The determination of the kinetic freeze-out temperature and its uncertainty is illustrated. 
    The lower horizontal axis belongs to the measured data points, the upper, temperature axis belongs to the model calculations.
    The value of the measured data point is projected onto the theoretical curve and the corresponding temperature is read off.}
    \label{fig:rho_temp}
\end{figure}

From the model described above, we determined the kinetic freeze-out temperature which would correspond to the available particle ratios from  heavy-ion data. 
The method how the temperature is extracted  is illustrated in Fig. \ref{fig:rho_temp}, and can be summarized as follows. 
From eqs.~(\ref{eq:evolvemu}) the chemical potentials are calculated as functions of temperature ($\mu_h(T)$).
The calculations were started from the initial values listed in Table \ref{tab:initial_values_for_calc}.
\begin{table}[t]
    \centering
    \caption{The initial temperature, chemical potentials and strangeness saturation parameters for the model calculations. In the case of ALICE, the values we utilized the parametrization published in Ref. \cite{Andronic:2017pug}., for STAR, the parameters are given in Ref. \cite{PhysRevC.96.044904}. For wider centrality bins (e.g. $0-10\%$ bin), we calculate the error-weighted-average from the centrality bins that cover the needed one. In the cases of tighter bins we use the same values in each bin which value was given for the covering one (e.g. $60-70\%$ and $70-80\%$ bins have the same value from the $60-80\%$ centrality bin).
    \label{tab:initial_values_for_calc}}
    \begin{tabular}{|c|c|c|c|c|}
        \hline
        \multicolumn{5}{|c|}{ALICE ($\sqrt{s_\textmd{NN}}=$ 2760 GeV)} \\
        \hline
        centrality & $T_\textmd{chem}$ [GeV] & $\mu_B$ [GeV] & $\mu_S$ [GeV] & $\gamma_S$ \\
        \hline
        $-$ & 0.159 & 0 & 0 & 1 \\
        \hline
        \multicolumn{5}{c}{} \\
        \hline
        \multicolumn{5}{|c|}{STAR ($\sqrt{s_\textmd{NN}}=$ 200 GeV)} \\
        \hline
        centrality & $T_\textmd{chem}$ [GeV] & $\mu_B$ [GeV] & $\mu_S$ [GeV] & $\gamma_S$ \\
        \hline
          0-5\% & 0.1643 & 0.0284 & 0.0056 & 0.93 \\
         5-10\% & 0.1635 & 0.0284 & 0.0050 & 0.95 \\
        10-20\% & 0.1624 & 0.0277 & 0.0059 & 0.94 \\
        20-30\% & 0.1639 & 0.0274 & 0.0064 & 0.90 \\
        30-40\% & 0.1616 & 0.0239 & 0.0060 & 0.90 \\
        40-50\% & 0.1623 & 0.0229 & 0.0058 & 0.84 \\
        50-60\% & 0.1623 & 0.0229 & 0.0058 & 0.84 \\
        60-70\% & 0.1613 & 0.0182 & 0.0054 & 0.76 \\
        70-80\% & 0.1613 & 0.0182 & 0.0054 & 0.76 \\
        \hline
         0-10\% & 0.1639 & 0.0284 & 0.0053 & 0.94 \\
        10-40\% & 0.1627 & 0.0264 & 0.0061 & 0.91 \\
        40-60\% & 0.1623 & 0.0229 & 0.0058 & 0.84 \\
        60-80\% & 0.1613 & 0.0182 & 0.0054 & 0.76 \\
        \hline
    \end{tabular}
\end{table}
In the case of STAR the $T_\textmd{chem}$, chemical freeze-out temperatures, differ slightly from what one would obtain from the parametrisation in \cite{Andronic:2017pug}.  
We decided to use the given values because they do not differ significantly from the one given by the parametrisation. 
On the other hand, we utilised the parametrisation for the higher energy \cite{PhysRevC.96.044904}.
The obtained $\mu_h(T)$  are employed to determine the number densities for each resonance species, as functions of $T$. 
Based on this, temperature dependence of the particle ratios can be predicted. 
This is illustrated in Fig.~\ref{fig:rho_temp}. on the example of $\rho^0/\pi$ ratio. 
The solid line in the Figure shows the calculated ratio as the function of temperature.

The calculations were done for every particle species for which measurements are available in the literature (see Tab. \ref{tab:alldata}.), 
and the temperature-dependent ratios were compared to the corresponding experimental data point(s). 
This comparison is illustrated in Fig. \ref{fig:rho_temp}. 
Notice the two horizontal scales on this plot:  
the bottom one corresponds to the data points, and the top scale is the temperature  of the model calculations. 
The kinetic freeze-out temperature can be determined by projecting the experimental values onto the theoretical curve, as illustrated with one data point in the Figure. 
The uncertainties can be determined in the same way.

It could happen that a data point or a measured uncertainty does not intercept with the calculated curve within the considered temperature range. 
In such cases we give an estimate, i.e., we do not give the kinetic freeze out temperature value with an uncertainty but we give the range where  the experimental uncertainty interval overlaps with the theoretical curve.


\section{Results}
\label{s:results}

The first ratio---$\rho^0/\pi$---was illustrated in Fig.~\ref{fig:rho_temp}. Qualitatively, we observe that as collisions become more central, the data points correspond to a lower temperature. The temperatures for all data points will be summarised later. 

\begin{figure}
    \centering
    \includegraphics[width=0.48\textwidth]{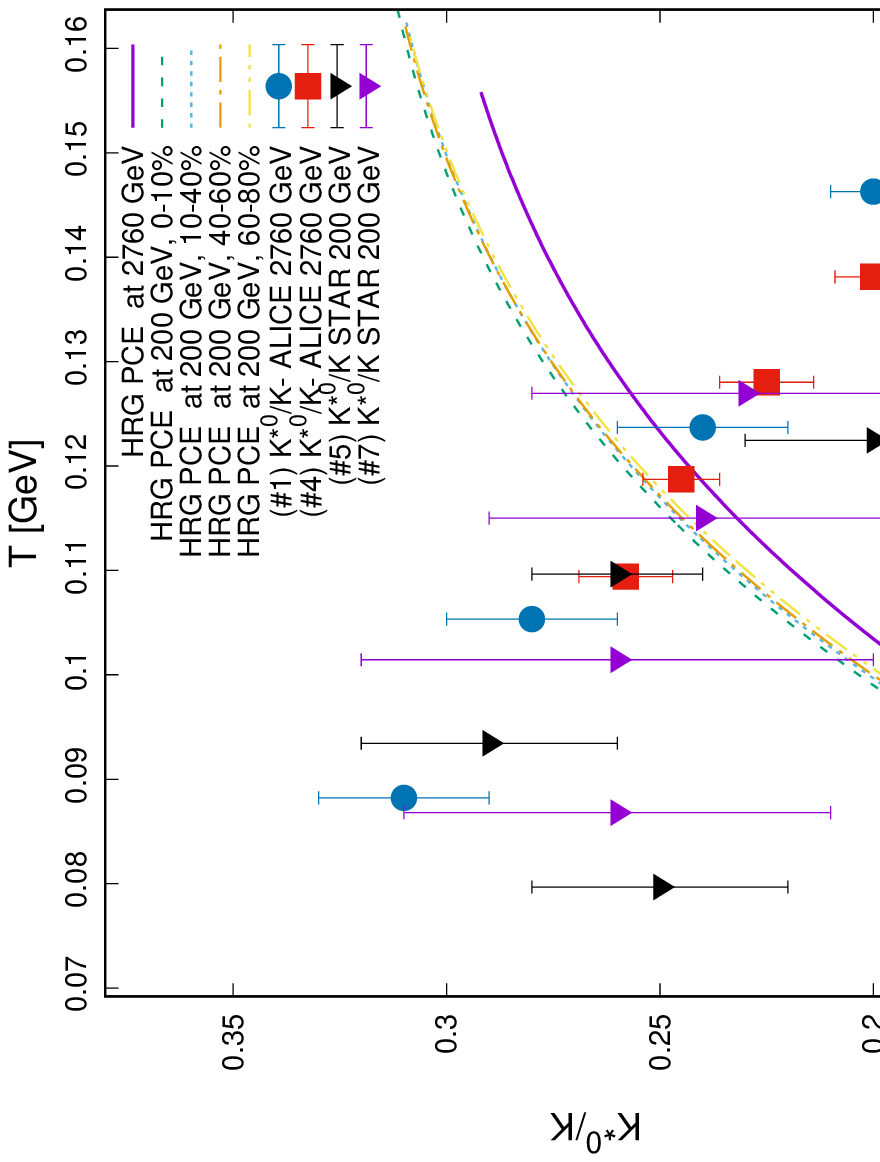}
    \caption{The $K^*/K$ ratios as functions of centrality (experimental data points) and temperature (theoretical curves). Solid curve stands for calculation for Pb+Pb collisions at $\sqrt{s_{NN}} = 2.76$~TeV (all centralities). Non-solid curves correspond to different centralities of Au+Au collisions at $\sqrt{s_{NN}} = 200$~GeV (and practically overlap). For the data points,
    the numbers in brackets refer to data set numbers from Table \ref{tab:alldata}.}
    \label{fig:ratiosKsK}
\end{figure}
\begin{figure}    
    \includegraphics[width=0.48\textwidth]{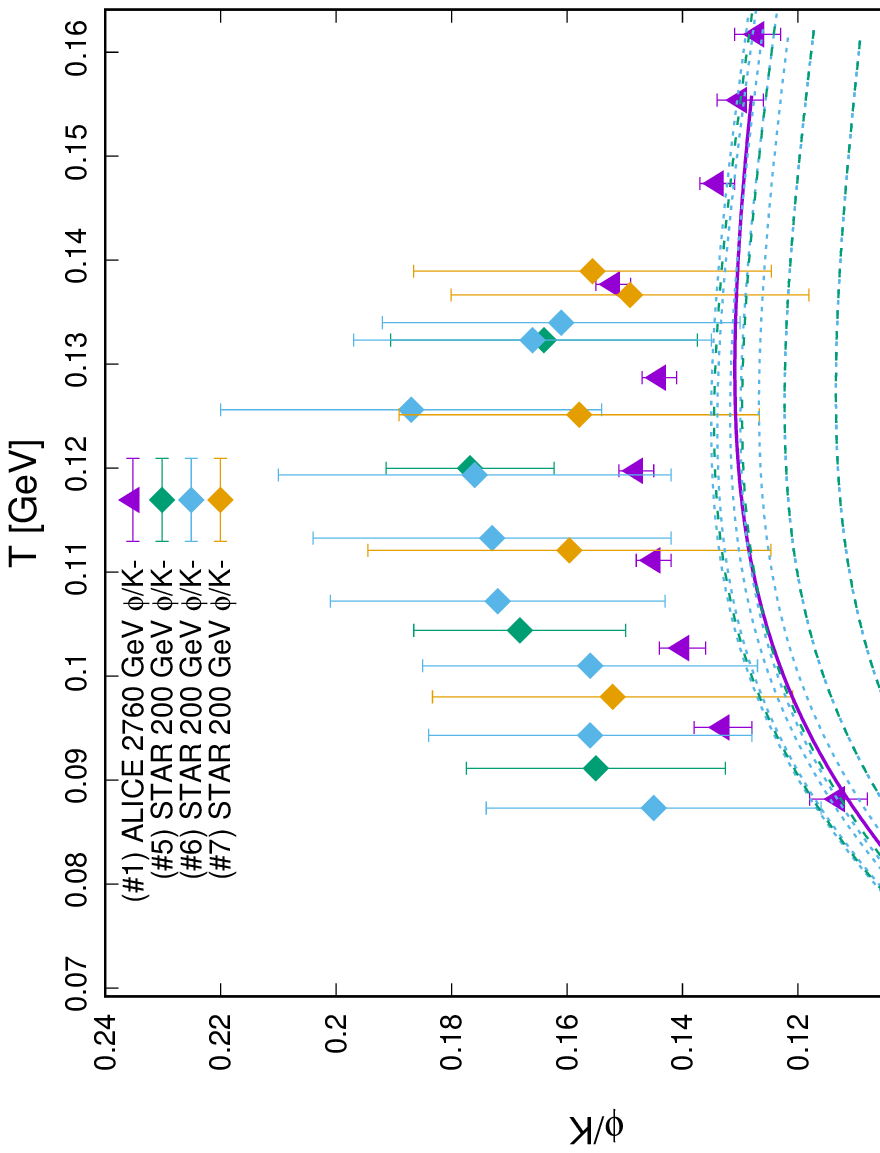}
    \caption{Same at Fig.~\ref{fig:ratiosKsK}, but for the $\phi/K$ ratios. Solid curve for Pb+Pb at $\sqrt{s_{NN}} = 2.76$~TeV (all centralities). Non-solid curves for Au+Au at $\sqrt{s_{NN}} = 200$~GeV from central  (top curves) to peripheral (bottom curves) collisions.
    For the data points,
    the numbers in brackets refer to data set numbers from Table \ref{tab:alldata}.}
    \label{fig:ratiosPhiK}
\end{figure}
\begin{figure}
    \includegraphics[width=0.48\textwidth]{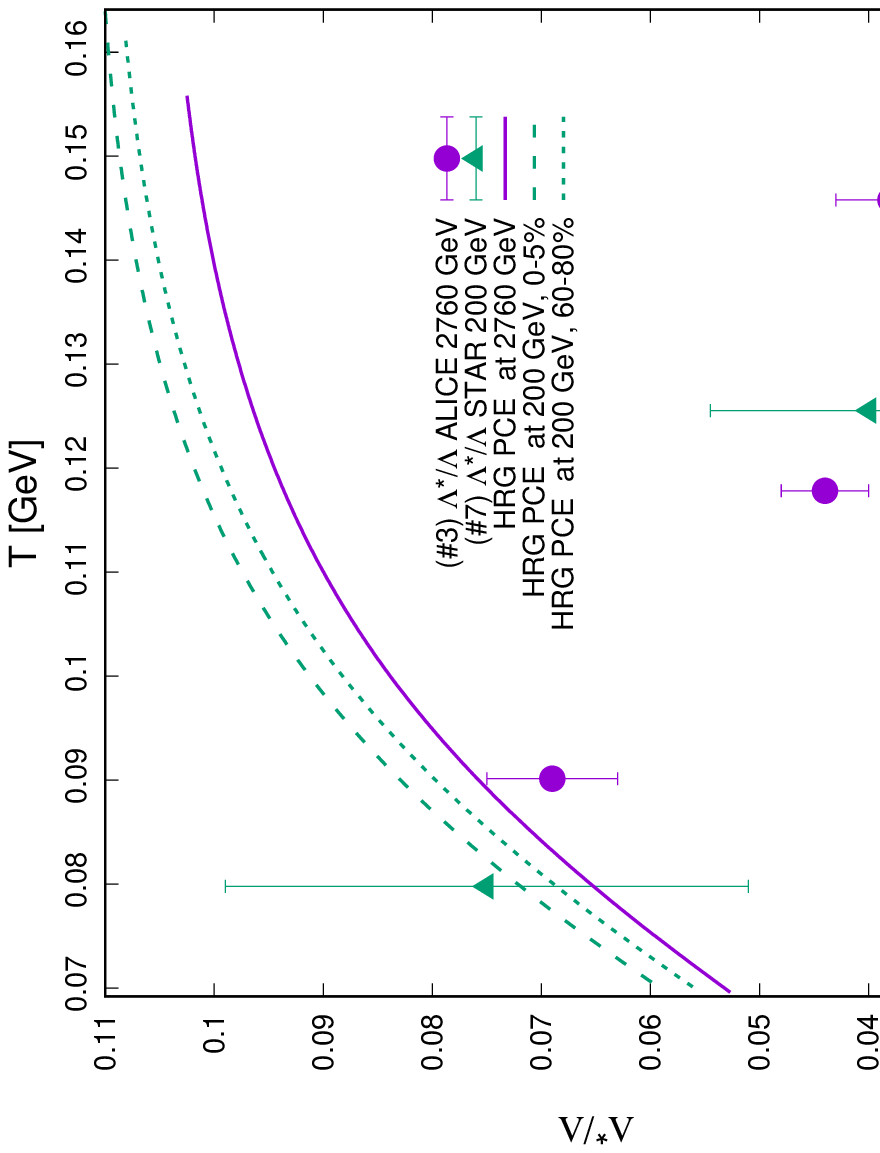}
    \caption{Same at Fig.~\ref{fig:ratiosKsK}, but for the $\Lambda^*/\Lambda$. Solid curve for Pb+Pb at $\sqrt{s_{NN}} = 2.76$~TeV (all centralities). Dashed and dotted lines for Au+Au collisions at $\sqrt{s_{NN}} = 200$~GeV, centrality 0--5\% and 60-80\%, respectively. For the data points,
    the numbers in brackets refer to data set numbers from Table \ref{tab:alldata}.}
    \label{fig:ratiosLsL}
\end{figure}
\begin{figure}
    \includegraphics[width=0.48\textwidth]{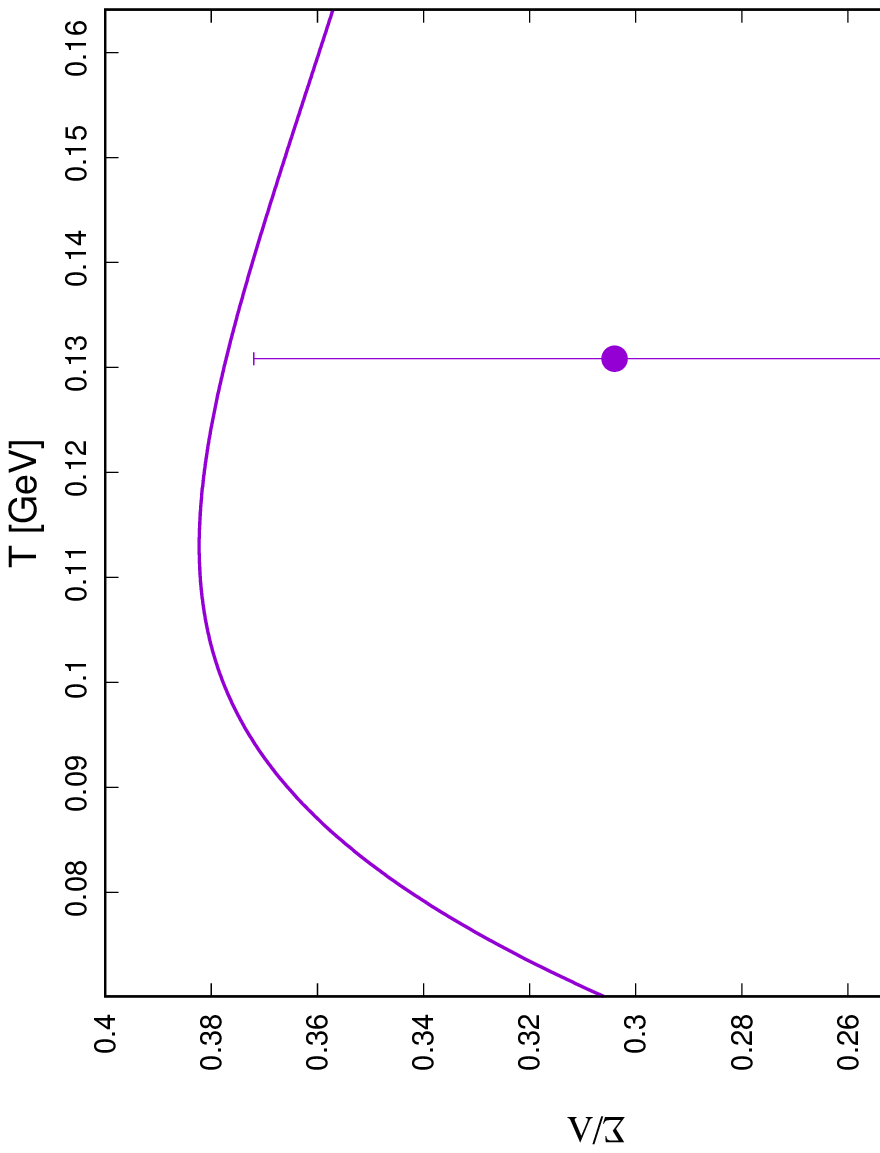}
    \caption{Same at Fig.~\ref{fig:ratiosKsK}, but for the
    $\Sigma^*/\Lambda$ ratio. Data point and calculation for Au+Au collisions at $\sqrt{s_{NN}} = 200$~GeV, centrality 0--5\%.}
    \label{fig:ratiosSL}
\end{figure}

Model calculations and other measured data points are superimposed in Figs. \ref{fig:ratiosKsK}-\ref{fig:ratiosSL}. 
In Fig.~\ref{fig:ratiosKsK}, we plot the data on $K^*/K$ ratio for different centralities
of Au+Au collisions at $\sNN = 200$~GeV as well as Pb+Pb collisions at $\sNN= 2.76$~TeV. At RHIC energies, the theoretical curves for different centralities are practically 
on top of each other, while there is always only one theoretical curve for ALICE, since the chemical potentials for all centralities are identical. 
The rough qualitative picture is again that more central data indicate lower kinetic freeze-out temperature. Nevertheless, the most peripheral ALICE data points do not match the theoretical curves 
anywhere and we can only find an overlap of the theoretical value with a fraction of the measured uncertainty interval. 

The disagreement of theory to experiment becomes most severe for the $\phi/K$ ratio, plotted in Fig.~\ref{fig:ratiosPhiK}.
Practically all measured data points are above the theoretical curve. 
In the present scheme, only the mesons from the pseudo-scalar octet 
are treated as stable. Hence, in spite of its rather long lifetime,
$\phi$ is treated as an unstable resonance. This means that it stays
always in equilibrium with its decay products, notably with $K$ and $\bar K$.
To stay consistently in framework of the PCE model, we do not modify 
this assumption. Note that the calculated ratio barely changes all the way down to the temperature $T=110$~MeV.
For the collisions at the LHC, the ratios in central collisions are 
reproduced by PCE calculations for a large interval of temperatures. 
However, the measured ratios for more peripheral collisions overshoot
the theoretical values. This also seems to be the case for all measured 
values at $\sNN = 200$~GeV. It indicates some non-equilibrium 
mechanism beyond the current PCE treatment, possibly the decoupling 
of $\phi$ from the tower of $K$ and $\bar K$ resonances.

An opposite situation appears for the $\Lambda^*/\Lambda$ ratio, see Fig.~\ref{fig:ratiosLsL}.
The overall trend seems similar as for the $K^*/K$: data for more central collisions 
appear to correspond to a lower temperature than those from peripheral collisions. 
However, the actual measured values are so low that the calculation would have to be 
brought to temperature below 70~MeV in order to reproduce central and mid-central 
data, 
but the fireball would have broken up before it would cool down so much. For two data 
points from $\sNN = 200$~GeV we can find an overlap with the theoretical curves only 
cthanks to the large error bars. 

There is only one data point measured for the $\Sigma^*/\Lambda$ ratio,
as seen in Fig.~\ref{fig:ratiosSL}. 
Even though the actual data point is below the theoretical curve, 
there is an overlap of the uncertainty interval with a portion of 
the curve, owing to the large measured error bars. 

\begin{figure}
    \centering
    \includegraphics[width=0.49\textwidth]{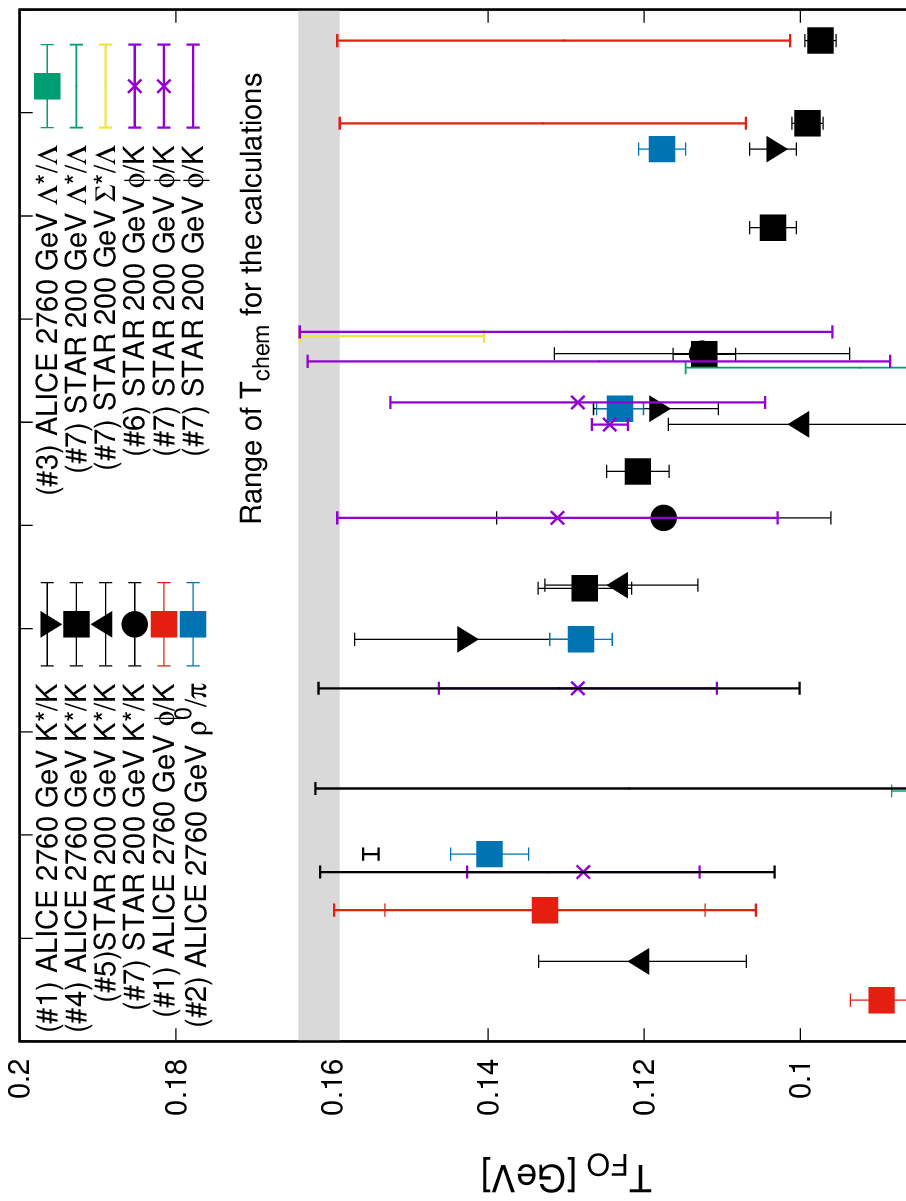}
    \caption{The freeze-out temperatures resulting from the measured values
    of resonance ratios. For the data points,
    the numbers in brackets refer to data set numbers from Table \ref{tab:alldata}.}
    \label{fig:temperatures}
\end{figure}
We summarize our results for the extracted temperatures in 
Fig.~\ref{fig:temperatures}, for all the ratios included in 
our analyses. In cases where we had the overlap of the measured 
value with the theory, we show the resulting temperature. If there 
was just a partial overlap with the uncertainty interval, we only 
show bars in the plot. Due to their large abundance, the dominant 
behaviour seems to be set by the $K^*/K$ data, which generally 
decrease when moving to more central collisions. 
The $\rho/\pi$ ratios 
seem to fall into the same temperature dependence, with some 
difference for the most central collisions. There, $\rho/\pi$
indicates a temperature higher by 15~MeV than the $K^*/K$ ratio. 
The ratios of $\phi/K$ and $\Lambda^*/\Lambda$ either fall out
of this temperature dependence or are connected with too large 
uncertainty intervals to make any reasonable conclusions.


\section{Conclusions}
\label{s:conc}

\begin{figure}
    \centering
    \includegraphics[width=0.49\textwidth]{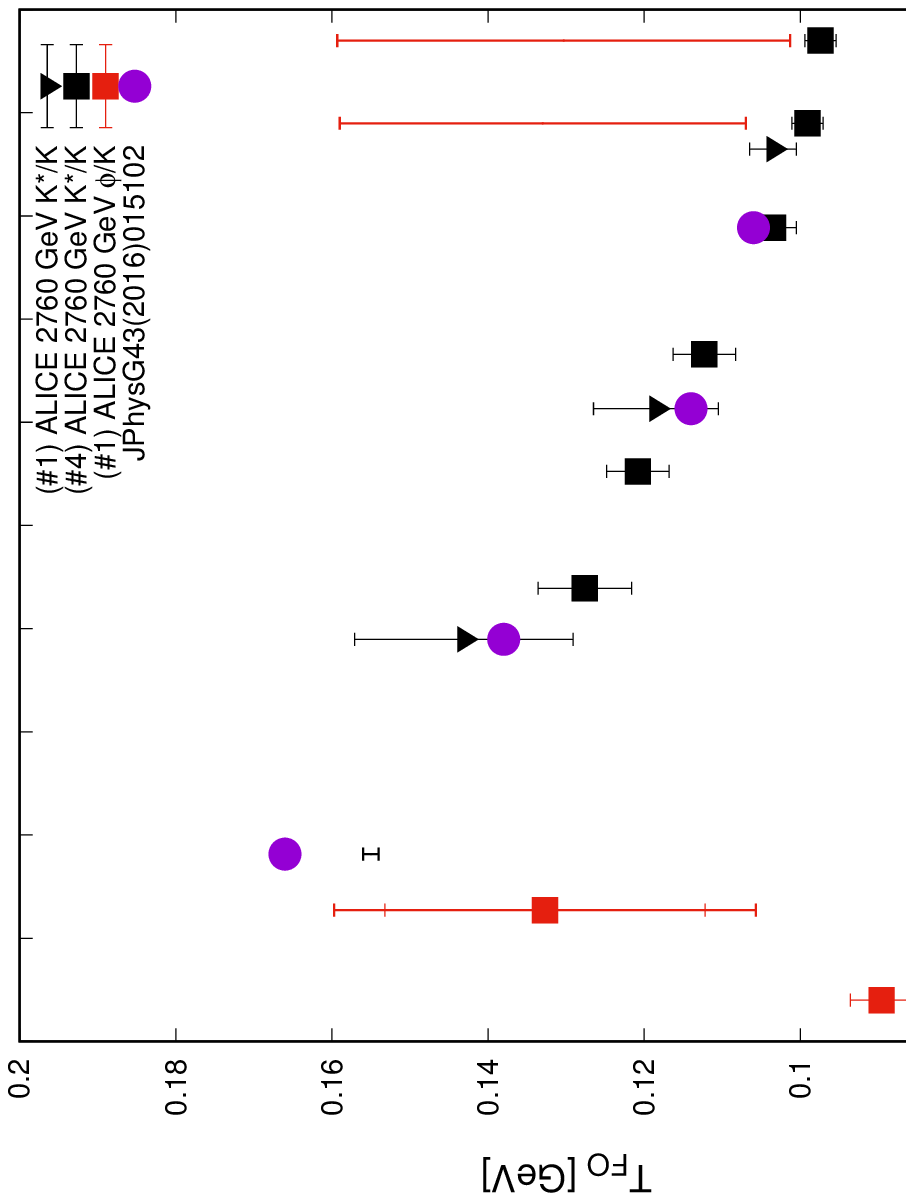}
    \caption{The freeze-out temperatures from the $K^*/K$ 
    ratios from Pb+Pb collisions at $\sNN=2.76$~TeV, compared with the kinetic 
    freeze-out temperatures obtained from fitting $p_t$ spectra of pions, 
    (anti)protons, and kaons \cite{Melo:2015wpa}. }
    \label{fig:kinetT}
\end{figure}

A part of the motivation for this study was to check if and how the resonance production
indicates the same freeze-out parameters as the single-particle $p_t$ spectra.
An analysis if the $p_t$ spectra, which lead to the kinetic freeze-out temperatures 
for different centralities of Pb+Pb collisions at $\sNN = 2.76$~TeV \cite{Melo:2015wpa}
yielded results that are consistent with those of $K^*/K$ ratios. This is 
plotted in Fig.~\ref{fig:kinetT}.

Nevertheless,  the partial chemical equilibrium is but one special scenario, which can be 
assumed for the evolution of the fireball after the chemical freeze-out. 

The disagreement of the $\phi/K$ ratio best points to the shortcomings of the model 
setup.
While $\phi$ is rather stable on the time scale of the hadronic fireball lifetime, 
the PCE treats it as unstable to such an extent, that it remains in equilibrium with its 
decay products. Accounting for $\phi$ as stable particle would possibly increase
its abundance.

This may be the strongest hint to the non-equilibrium 
behaviour of higher-mass states and possible rescattering of 
their daughter particles \cite{Knospe:2015nva}. 
Such an mechanism  would also include the  scenario
with a short hadron phase \cite{Mazeliauskas:2019ifr}.

Another shortcoming of the PCE model is the assumption of isentropic evolution. It remains 
to be studied in the future, how this assumption can be relaxed and what impact 
on the results it would have.

\acknowledgments

SL is grateful of the support of Hungarian National E\"otv\"os Grant established by the Hungarian Government.
BT ackowledges the support by VEGA grant No 1/0521/22.

\pagebreak

\bibliography{sample}

\end{document}